# Intermitotic timing and motility patterns in the cell division of the diatom *Seminavis robusta*

Jonas Ziebarth, Thomas Fuhrmann-Lieker
Center for Interdisciplinary Nanostructure Science and Technology (CINSaT),
University of Kassel, Heinrich-Plett-Str. 40, 34109 Kassel, Germany
th.fuhrmann@uni-kassel.de

**Abstract**
Many diatoms follow a size diminuation - size restoration cycle in their vegetative phase, leading to daughter cells that differ in size. For the diatom *Seminavis robusta*, we investigated by cell tracking over several generations whether the size difference reflects also in different intermitotic times or in the mobility of the cells. A tracking setup and machine-learning based detection algorithm was developed that revealed no significant difference in intermitotic times, a weak coupling to the day-night cycle, and a higher motility of the hypothecal, smaller daughter cell.

**Introduction**
Diatoms (Bacillariophyta) as one of the major contributors to photosynthesis in aquatic systems worldwide are characterized by an unique life cycle, coined the „Diatom Sex Clock" by W. Lewis.[1] Since the cell wall consists of two overlapping porous but rigid silica shells, named epitheca and hypotheca, growth in diameter (centric diatoms) or length (pennate diatoms) is not possible. Therefore in the vegetative phase, mitosis has to provide special features for cell division. Instead of simple cytokinesis, the membrane is first invaginated, separating both daughter cells within a common cell wall, then in each half a new smaller hypotheca fitting into the parental theca is synthesized, until finally the daughter cells separate. Whereas one of the daughter cells retains the original size of the mother, the other one is smaller, such that in a population the average cell size is decreasing from generation to generation until sexual reproduction yields new large initial cells. Notably there are some exceptions from this size reduction-restoration cycle such as the genetic model diatoms *Phaeodactylum tricornutum* and *Thalassiosira pseudonana*,[2] and other centric diatoms[3], but typically species follow this so-called MacDonald-Pfitzer rule, including the pennate diatom *Seminavis robusta* used in our investigations.

The different size of epithecal and hypothecal daughters – sometimes called older and younger daughter because of the age of their epitheca – introduces an asymmetry in the cell cycle. A delay in intermitotic time, i.e. a longer subsequent time to the next division, of the smaller, hypothecal daughter was observed by O. Müller for the chain forming *Ellerbeckia* (formerly *Melosira*) *arenaria* as early as in 1883.[4] More recently, Laney *et al.* just reported the opposite for *Ditylum brightwellii*.[5]

A delay in intermitotic times has consequences for models of the size-reduction-restoration cycle.[6] Whereas the size distribution of the normal MacDonald-Pfitzer rule follows Pascal's triangle and leads to an exponential growth of the population, a delay of one generation for one of the daughters shifts the development of the cell number to a Fibonacci sequence.[7,8,9]

In order to decide about a possible mismatch in the intermitotic timing of the daughter cells, and for observing any other kind of asymmetry in the cell division process of the proposed model diatom for the life cycle *Seminavis robusta*, we performed statistical tracking experiments for this species over several generations. *S. robusta* is a benthic diatom that is able to glide on the substrate by excretions from its raphe (Fig. 1), the slit for the segregation of adhesive strands presumably connected with actin-myosin motors.[10] The special challenge of these tracking experiments is the continuous generation of new cells which have to be included, the assignment of epithecal and hypothecal daughters, respecively, and the sufficient spatial and temporal resolution for a long timespan.

**Setup and tracking analysis**

*Seminavis robusta* was obtained from the Belgian Coordinated Collection of Microorganism in Ghent as strain PONTON26 (DCG0457, mating type "+"). For experiments, Si-f/2 medium[11] was inoculated with stock culture from the same medium in a 30 ml Petri dish, and the cultures were subjected to a 12:12 LD entrainment with white light at 25 µmol m$^{-2}$ s$^{-1}$ in a Sanyo MIR 153 incubator for at least 5 days. Then the cultures were transferred to a standard 384-well plate with ~20 cells in the observation well and monitored under a wide-field microscope (Leica DMIL with objective HI PLAN 4x/0.10 PH0), the field of view covering a well almost completely. Two light sources are used: A near-IR LED at 770 nm was used as the illumination source for tracking, whereas a broad-band LED light source (Aqua Illumination Prime 16HD) continued the entrainment at 25 µmol m$^{-2}$ s$^{-1}$ for additional 5 days. A 768 nm band pass filter was inserted between objective and camera (DFK 33UX226). 4K pictures were recorded every 10 s using IC Capture (The Imaging Source Europe, Germany). The experiments started at the onset of the dark phase and ran for at least one week.

A custom Python code was written for the analysis of tracks. First, semantic segmentation based on the modified UNet code from Lee et. al.[12] was applied. In this, pixels are labeled as belonging to diatom cells, aggregates/debris, or background, and subsequently merged into connected objects. Supervised training with automatically generated test data sets of 100 microscopy images was employed. The miss rate of cells was MR=30%, the coefficient of fragmentation (mean number of labels per ground-truth object) CF=0.7, and the detection coefficient (number ratio of detected objects and ground-truth objects) DC=1.1. Only 5% of all detected particles were classified as aggregates/debris.

Second, detected cells' contours were simplified by fitting boundary boxes using the OpenCV library,[13] including orientational data. Thus, each object is characterized by a location ($x$, $y$), bounding box lengths ($a$, $b$), and a rotational angle $o$.

Third, for tracking in subsequent images, Kalman filters were used as base for the tracking which adapts the SORT (Simple Online and Realtime Tracking) algorithm.[14] The algorithm was modified by a constant acceleration model[15] and extended to track orientation, dividing events and heritage. The Kalman filter predicts the location of the object based on its previous movements. Here, the five state variables and their time derivatives ($\dot{x}, \dot{y}, \dot{a}, \dot{b}, \dot{o}, \ddot{x}, \ddot{y}$) were used as state vector for the filter. In addition to cells in a rest position, three principal patterns of motion behavior are observed with *Seminavis robusta*: forward movement on circular arcs, abrupt stopping, and reversal of direction.[16] To account for these possibilites, an assignment was calculated first by comparing to the cells's last position, then the other three options of location or velocity shift were taken into account (with Intersection over Union metrics). Irregular movements were accounted for by additional Euclidean distance metrics. By inflating the covariance matrix with a factor of 1.01, the Kalman filter is deprived partially from its memory so that it becomes more responsive to updates.[17]

If a bounding box remained unmatched at the end, a new Kalman filter was assigned. Thus, if a filter matched with two objects, it was concluded that the cell has divided, and two new duplicates, representing the daughter cells, were created and continue instead of the previous filter which was then removed. As intermitotic time, the time between such events was measured. Since not all tracks and lineages could be completed automatically, a manual revision step involving an inspection of the videos and fusing tracklets to complete tracks was applied in order to complete the lineages as much as possible. Estimated 22% of the detections were missing (recall value 0.78), and some lineage trees remained incomplete.

The most difficult task that had to be addressed is the identification of epitheca and hypotheca. For each division event, an assignment had to be made in order to be able to compare both with respect to their internal timing and motion behaviour. This exceeds the standard tracking schemes, so we will describe in detail how this problem was solved. To the best of our knowledge, there has been no similar attempt in cell tracking yet.

The difference in cell size between the two daughters is too small to be a reliable measure. But the asymmetric structure of the thecae and details of the cell division help here. The raphe is located at the ventral side of the cells. In *Seminavis robusta*, it is not centered, and the theca is flattened around it, so that during motion the cell is slightly tilted. In the microscope, the cells obtain their characteristic banana-shaped form. If the cells rest on their girdle band, which happens in rest and upon divion, they look symmetrical. In division, the two thecae fan out, and thus the two old thecae can be identified by the outer positions (see also Fig. 1b, in which theca are stained partially with Rhodamine 110 according to the procedure in ref.[18]). Since the cells do not rotate in this process, complete tracking of the orientation of the long axes allows an assignment of epi- and hypothecal daughter after the second division.

The derived trajectories, coloured differently for the hypothecal and epithecal daughters, were visually inspected at the graphical user interface (created with Tkinter), and false assignments were corrected if necessary (Fig. 2). A total of 3 tracking experiments with tracklets from ~360 cells were performed.

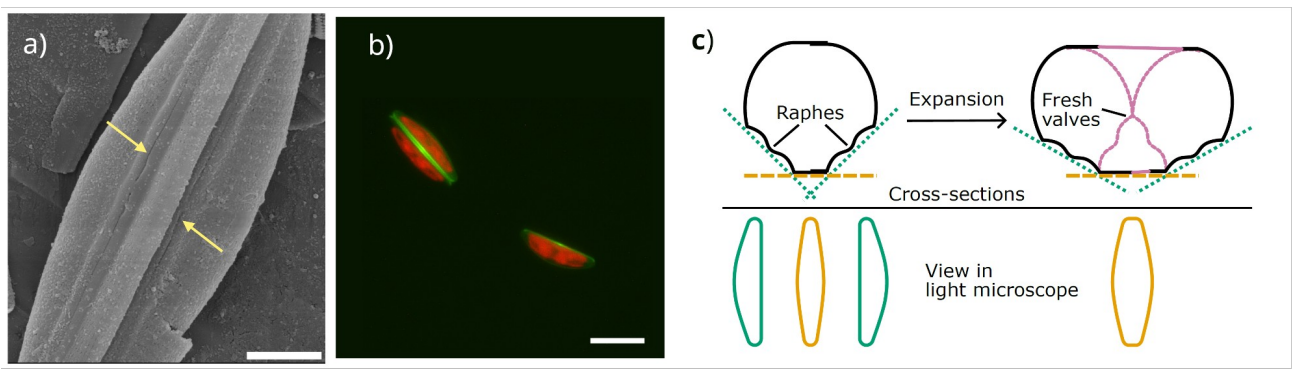

Fig. 1 Morphology of the Seminavis robusta theca. a) Scanning electron microscopy image of a cell showing the raphe of epitheca and hypotheca (arrow). Scale bar 4 µm. b) Position at cell division and in rest phase, imaged by fluorescence microscopy. Red are the plastids, green the stained newly synthesized theca parts Scale bar 20 µm. c) Cross-section and corresponding microscopic views for different tilt angles. Substrate planes are indicated by dotted and dashed lines.

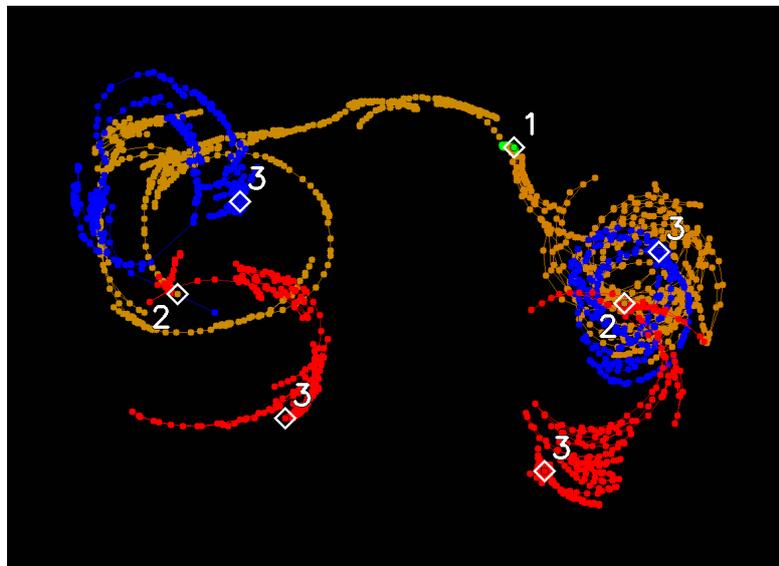

Fig. 2 Example trajectory. Tracklets can be assigned to epithecal (blue) and hypothecal daughter (red) after the first division. Division events are marked with numbers for each generation.

## Results
### 1. Intermitotic times
In 220 cases, a complete cell cycle was measured and an intermitotic time defined as described above was calculated. For this subset the mean intermitotic time was 17.14 h with a standard deviation of 4.59 h. It was possible to construct lineage trees with pairs of sisters and cousins for which assignment of epithecal or hypothecal offspring was possible after the second generation. Some trees remained incomplete due to missing tracklets with possible orientation switches. Fig. 3 displays a typical generation tree arising from a tracking experiment. Visual inspection already shows that the intermitotic times do not vary strongly between siblings of the same generation, but from one generation to the next. The influence of zeitgeber time under LD conditions is discussed in a later section.

A statistic comparison of the median intermitotic times of epithecal and hypothecal daughters is displayed as a boxplot in Fig. 4. An unpaired *t*-test showed no significant difference between hypothecal and epithecal daughters as ensembles ($p$=0.623, $N_1$=$N_2$=48). Also if sisters are compared directly, no significant difference is detected (paired *t*-test, $p$=0.458, $N_1$=$N_2$=48).

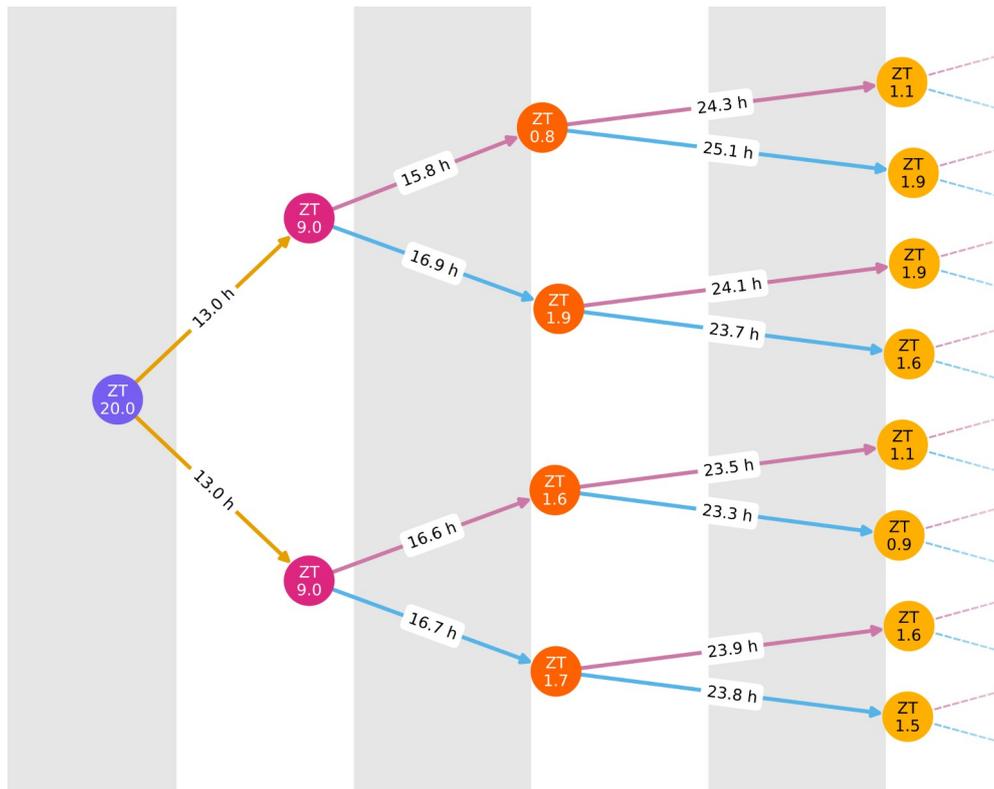

Fig. 3 Lineage tree of a typical tracking experiment. Epithecal (hypothecal) daughters are marked in blue (red). LD pattern and zeitgeber time (ZT) is indicated, as well as intermitotic times.

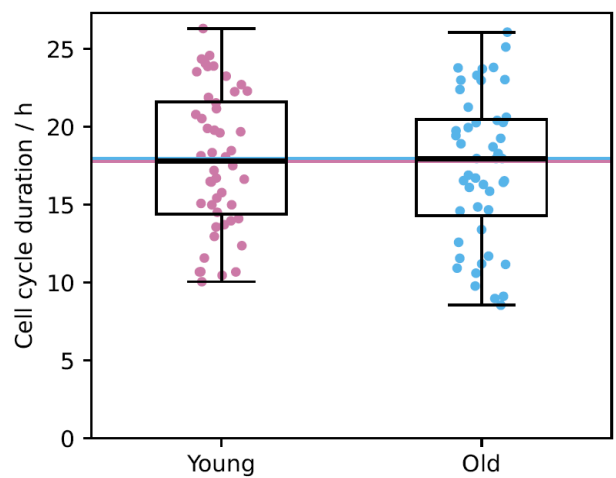

Fig. 4 Comparison of median intermitotic times between hypothecal (left) and epithecal (right) daughters (no. of cells $N_1$=$N_2$=48). Medians are indicated as horizontal lines. Box limits represent the quartiles and whiskers are extended to cover the total range of observed times.

## 2. Motility

The tracking experiments allow the calculation of motility parameters after cell division events, i.e. traveled distance and speed. Again, epithecal and hypothecal daughters are compared. The cells move along arc-shaped trajectories. Therefore in Fig. 5 both the average traveled distance along the trajectory as absolute values irrespective of the direction, as well as the average distance from the place of division is shown. Here, a significant difference between the daughters is detected: the smaller, hypothecal daughters dislocate more quickly along their trajectories (Mann-Whitney *U*-test: p=0.005, Wilcoxon *T*-test: $p=0.012$, $N_1=N_2=48$). Because of the higher standard deviation (around 50%), the difference in the distance from origin is less significant but detectable (*U*-test: $p=0.022$, *T*-test: $p=0.054$). Hypothecal daughters reached almost twice the distance than their epithecal sisters. At the beginning the curves exhibit linear slopes indicating a ballistic law with constant velocity but subsequently the mean velocity drops down (Fig. 5a. shows the travelled distance in both directions on the arc without sign, so the velocity can be directly extracted as slope). Thus also in 2D (Fig. 5b), the diffusion-like behaviour is flattened for longer times.

The difference in motility is also reflected in the histograms of velocities that show the speed distributions separately for epi- and hypothecal daughters. Since the distribution is asymmetric with a high contribution of slow moving phases, a *U*-test was performed. Again the difference is significant ($p<0.001$, analyzed frame differences $n_1=8072$, $n_2=4489$). Hypothecal daughters show a mean of 2.54 µm/s (median 2.20 µm/s), epithecal daughters 2.00 µm/s (median 1.77 µm/s).

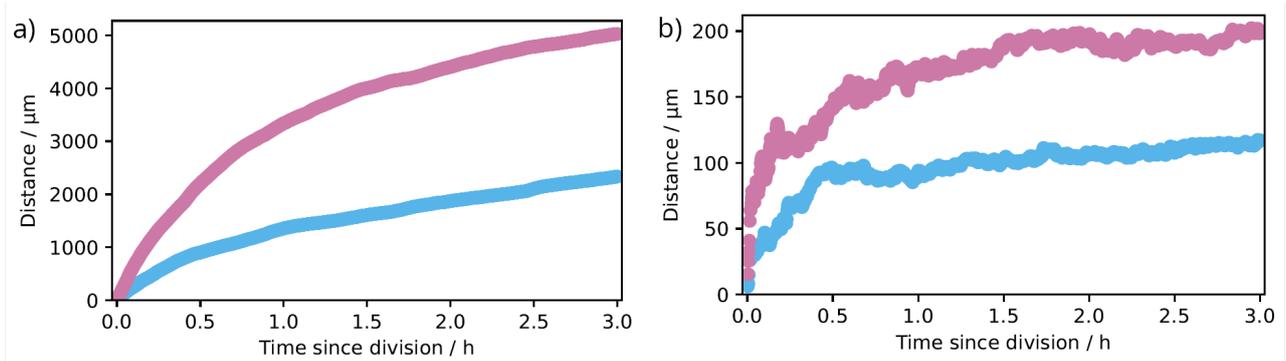

Fig. 5 a) Travelled distance along trajectories and b) distance from origin for hypothecal (pink, upper line) and epithecal daughters (blue, lower line)

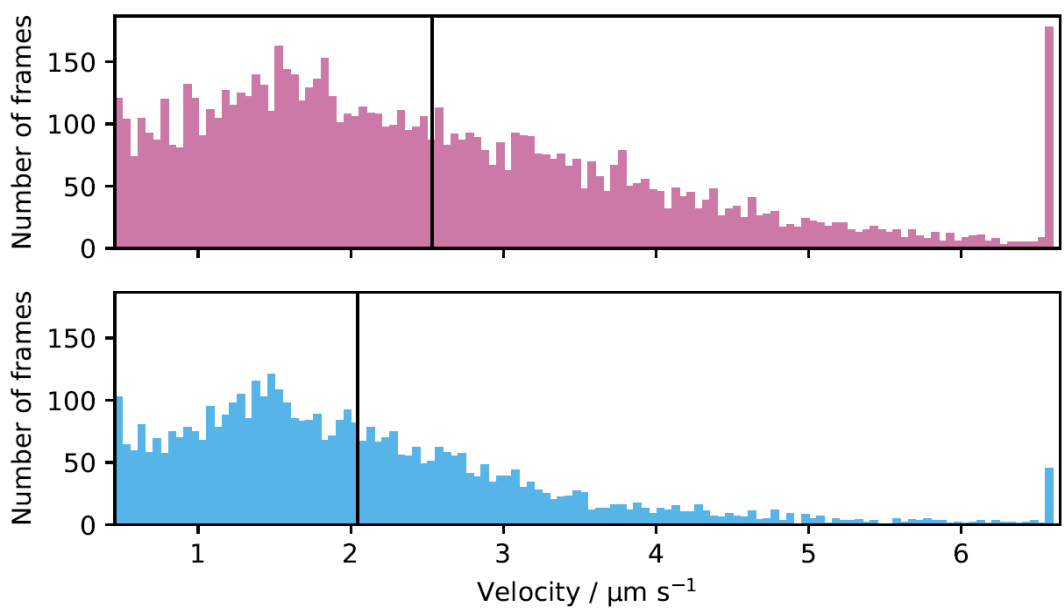

Fig. 6 Velocity distribution for hypothecal (pink, upper plot) and epithecal (blue, lower plot) daughters displayed as histogram. Higher velcocities as the plotted range are accumulated in the last bar. Mean values are indicated.

## 3. Coupling to *zeitgeber*

Both intermitotic time and motility may be also influenced by the external influence of light under the culture conditions. This is reasonable since for the related species *Phaeodactylum tricornutum*, a light-dependent checkpoint in the G1 phase of the cell cycle was found, activating a diatom-specific cyclin for proceeding towards the S phase which then takes place preferably in the night where DNA replication is protected from UV light.[19] This allows synchronization after a prolonged dark period and was demonstrated also for other species, including *Seminavis robusta*.[20] For lineage tracking under regular light-dark cycles, this is one explanation why cell cycle durations between sisters tend to be similar unless there is an internal difference in timing, but it predicts also that cycle lengths between cousins may be more similar than between mother and daughter cells. This model, comprising an internal clock coupled with an external trigger, is referred to as „kicked cell cycle".[21,22]

Thus we plot intermitotic time against the zeitgeber time in which the cell cycle started. Since in our tracking experiments the division event is detected by the first dislocation, the onset of motility and cell cycle can not be distinguished. The general picture is given in Fig. 7a) for 215 cells. Cycles shorter than 4 h are omitted. Individual cells are marked as dots. The density of dots showed that divisions occur all the day and night, but there is a maximum at the beginning of the day. This correlates with the observation that the maximum velocity of cells is detected at this time (Fig. 7b). On average, cycles starting in the last night hours are longer, the ones starting in the last day hours shorter, but the spread is high.

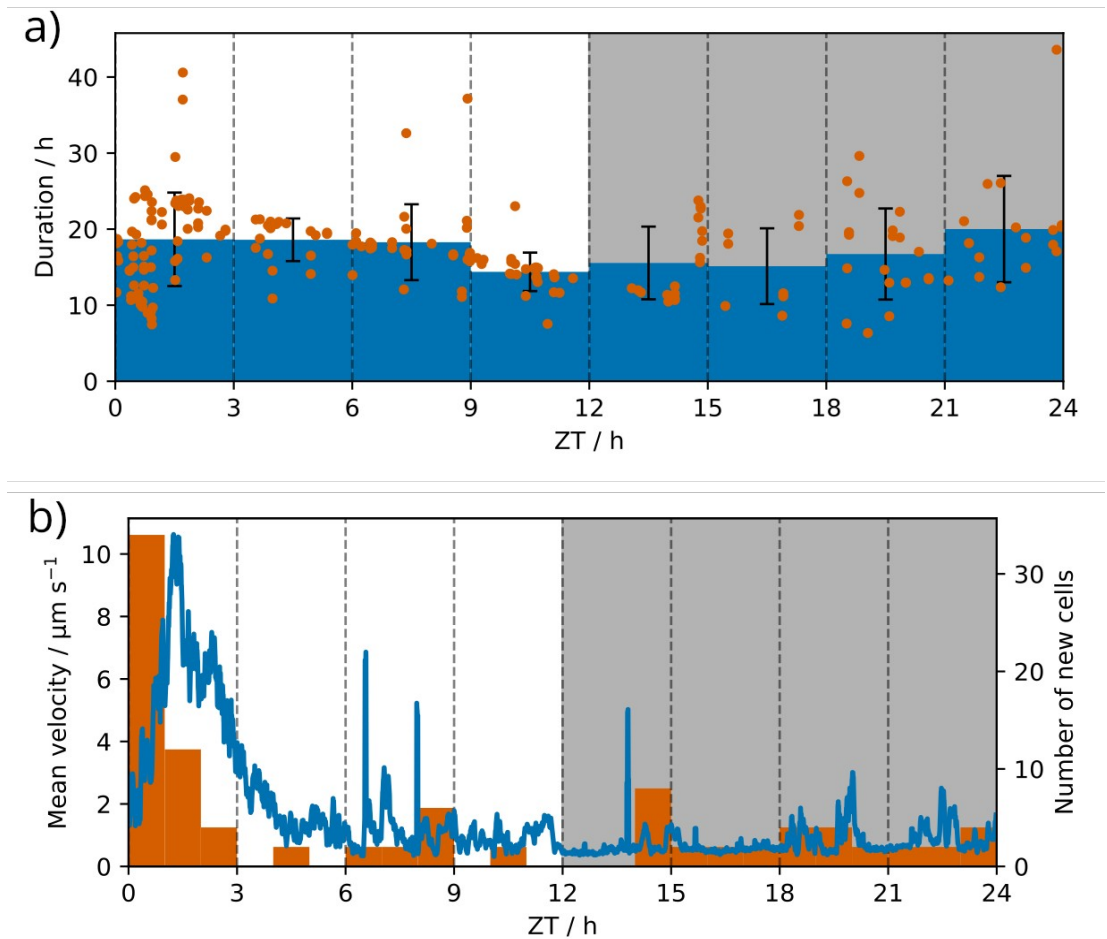

Fig. 7 a) Cell cycle duration after division at different ZT. The bars represent binned data in 3 h intervals with standard deviation in black. b) Mean velocity vs. zeitgeber time (blue curve). The orange bars represent number of new cells binned in 1 h intervals.

In order to find indications for a kicked cell cycle, we correlate intermitotic times of different relatives. The Spearman correlation table is given in Table 1. We found a significant dependence on the degree of relationsship ($p<0.05$). The highest correlation is between siblings (sister cells) with a positive Spearman correlation of +0.61 ($N$=83), followed by cousins (+0.46, $N$=56), whereas the parent-child correlation is slightly negative (-0.21, $N$=129). So indeed, a kicked cell cycle would explain the experimental findings.

Table 1: Correlation if cell cycle duration between relatives.
$\rho$: Spearman's coefficient, $N$: number of pairs

|  | $\rho$ | $N$ |
| --- | --- | --- |
| Parent-child | -0.21 | 129 |
| Siblings | +0.61 | 83 |
| Cousins | +0.46 | 56 |

**Further Discussion and Conclusion**

We were able to track diatom cells of *S. robusta* between several cell divison events with a distinction of epi- and hypothecal offspring, a feature that cannot be done by standard tracking software. Thus it was possible to correlate cell cycle duration and motility pattern between relatives of different degree.

In contrast to reports on *Ellerbeckia arenaria* and *Ditylum brightwellii*, we did not find a significant difference in intermitotic times between two sisters. Since *E. arenaria* is a freshwater chain-forming centric diatom, *D. brightwellii* a marine planktonic centric diatom, and *S. robusta* a marine benthic pennate diatom, this reflects the high variety of cell cycle timing patterns within the diatom clade. For a description of the life cyle, the sex clock, it is clear, however, that *S. robusta* follows a binomial growth in cell number and not a Fibonacci pattern which would be appropriate if one daughter showed a delay in cell division by one generation.

In motility on the other hand, a significant difference is detected. Generally, one daughter starts to move actively away from the place of division (as can be seen in the tracking movies), and it is the smaller hypothecal daughter that statistically moves farther away and with higher velocity. That the two daughters move apart may be due to competition for nutrients. That the younger, hypothecal daughter is more mobile, cannot be clearly attributed to the size since the difference is not so pronounced.

The cell cycle duration lies in the order of one day, and there is a preference for cell separation in the early morning hours. Due to the coupling with the day-night cycle, a higher correlation of cell cycle durations between cousins than between generations was found, according to the kicked cell cycle model that is applied also for other organisms.

**Acknowledgements**
This work was partly funded by the Deutsche Forschungsgemeinschaft, GRK2749, contract no. 448909517 (Biological Clocks on Multiple Timescales).

**Author contributions**
J.Z. built the setup, performed the experiments, wrote the analysis software, and did the analysis.
T.F. devised the topic, supervised the work, and wrote the text of the paper.

**Data availability**
Original tracks and data sets are deposited at the repository of the University of Kassel DaKS.

**Competing Interests**
There are no competing financial or non-financial interests.